\documentclass[12pt]{article}

\input epsf
\usepackage{amssymb,amsmath,wrapfig}
\usepackage[matrix,arrow,curve]{xy}
\usepackage{graphicx}
\usepackage{slashed}
\usepackage{calligra}

\DeclareMathAlphabet{\mathcalligra}{T1}{calligra}{m}{n}
\DeclareFontShape{T1}{calligra}{m}{n}{<->s*[2.2]callig15}{}

\makeatletter
\@addtoreset{equation}{section}
\makeatother

\def\del {\partial}

\def\del {\partial}

\newcommand{\scripty}[1]{\ensuremath{\mathcalligra{#1}}}

\def\del          {\partial}

\def\tr           {\mathop{\rm Tr}}

\def\half{{\frac12}}

\def\sqr#1#2{{\vcenter{\vbox{\hrule height.#2pt
 \hbox{\vrule width.#2pt height#1pt \kern#1pt \vrule width.#2pt}\hrule
 height.#2pt}}}}

\def\beq {\begin{equation}}
\def\bea {\begin{eqnarray}}
\def\eeq {\end{equation}}
\def\eea {\end{eqnarray}}

\def\del{\partial}

\def\CN {{\cal N}}

\def\half{\frac{1}{2}}

\newcommand{\Tr}{{\rm Tr\,}}

\begin{document}

            \begin{titlepage}

            \begin{center}

            \vskip .3in \noindent

            {\large \bf{The Exact Superconformal R-Symmetry
            Extremizes\ 
            $Z$}}

            \bigskip

                Daniel~L.~Jafferis$^1$

            \bigskip

            $^1$ School of Natural Sciences,
        Institute for Advanced Study, Princeton, NJ 08540, USA\\
        \vspace{.1cm}

            \vskip .5in
            {\bf Abstract }
            \vskip .1in

            \end{center}

            \noindent

The three sphere partition function, $Z$,  of three dimensional
theories with four supercharges and an $R$-symmetry is computed
using localization, resulting in a matrix integral over the Cartan
of the gauge group. There is a family of couplings to the curved
background, parameterized by a choice of $R$-charge, such that
supersymmetry is preserved; $Z$ is a function of those parameters.
The magnitude of the result is shown to be extremized for the
superconformal $R$-charge of the infrared conformal field theory,
in the absence of mixing of the $R$-symmetry with accidental
symmetries. This exactly determines the IR superconformal
$R$-charge.

            \vfill
            \eject


            \end{titlepage}

\begin{section}{Introduction}

This paper contains evidence for two results. The first is that
the $S^3$ partition function of a theory with four supercharges
and an $R$-symmetry can be computed using localization. The
partition function depends on supersymmetry preserving couplings
to curvature, which are parameterized by a choice of $R$-charge.
These differ by abelian non-$R$ flavor symmetries. The second
result is that the magnitude of that partition function is
extremized for the superconformal $R$-charge of the 3d infrared
conformal field theory.

Under the assumption that the $R$-symmetry does not mix with
accidental flavor symmetries, the partition function on the sphere
can be computed in a weakly coupled UV theory. The results of this
paper then lead to explicit exact formulas for the superconformal
$R$-charge in the IR.

There has recently been considerable interest in the partition
function of supersymmetric three dimensional theories on $S^3$.
This quantity is extensive in the number of degrees of freedom in
the system, yet it can be calculated exactly using
supersymmetry. Moreover the resulting answers can be written in closed form, unlike the perhaps richer invariants given by the superconformal index on $S^2 \times S^1$ studied in \cite{Kim}.

The technique of localization of supersymmetric partition
functions involves the addition of a $Q$-exact operator to the
action, which does not affect the path integral, but renders the
1-loop approximation exact. This was applied to the
infrared-finite problem of gauge theories on spheres by Pestun
\cite{Pestun}, who obtained exact results for ${\cal N}=2$ 4d
theories on $S^4$.  The study of the three dimensional case was
initiated by Kapustin, Willett, and Yaakov \cite{KWY1}, for
theories with ${\cal N}=2$ supersymmetry and no anomalous
dimensions for chiral operators.

In practice, this means that it has been used in 3d theories with
${\cal N} =3 $ or more supersymmetry, since their nonabelian
$R$-charges cannot receive quantum corrections. In this work, I
will generalize the localization argument to any ${\cal N}=2$
three dimensional field theory with an $R$-symmetry.

In particular, I will explain how the $S^3$ partition function of
an ${\cal N}=2$ superconformal field theory can be computed
exactly at 1-loop from a localized UV Lagrangian, given the input
of the $R$-charge that controls the dimensions of operators in the
infrared CFT.

The full path integral on the sphere localizes to a matrix model,
which may be solved in the 't Hooft limit using large $N$
techniques. This method has been put to impressive use by Drukker,
Marino, and Putrov \cite{DMP} (see also \cite{HKPT, SMP}), who found the famous $N^{3/2}$
scaling of the entropy of multiple M2 branes directly in the IR
${\cal N}=6$ conformal field theory \cite{ABJM} on $N$ M2 branes.

Recall that in three dimensions, there are no anomalies in the
trace of the stress energy tensor, and thus no obvious analogues
of the $c$ and $a$ anomaly coefficients of four dimensional
theories. The quantity $a$ is the coefficient of the Euler
density, and $c$ is the coefficient of the square of the Weyl
tensor.

The $a$ function can be expressed simply \cite{AEFJ, AFGJ} in
terms of the superconformal $R$-charge in 4d ${\cal N}=1$
theories. Moreover, it was shown by Intriligator and Wecht that
$a$ is maximized as a function of a trial $R$-charge, thus
determining the exact superconformal one \cite{amax}. In this
work, evidence will be given that,  in 3d theories with ${\cal
N}=2$ supersymmetry, the sphere partition function, $Z$, plays a
very similar role. An intriguing and seemingly different proposal
for a quantity that behaves monotonically along renormalization
group flow is the entanglement entropy defined and
holographically studied by \cite{MS, MS2}.\footnote{In
\cite{CHM}, which appeared after the first version of this paper,  it is shown that these two quantities are in fact equivalent.}

 The partition function of a four dimensional theory
on $S^4$ suffers from logarithmic divergences. Thus the exact
results of Pestun \cite{Pestun} have usually been applied to
ratios of such partition functions, or VEVs of BPS operators. The coefficient of the logarithmic divergence in the free energy is proportional to the $a$ function multiplying the Euler density in the trace anomaly (the Weyl curvature vanishes on $S^4$, so $c$ does not appear). This was one of the original motivations of Cardy \cite{Cardy} in proposing that $a$ was should be decrease in the IR.

In contrast, for  three dimensional conformal theories, the
partition function on the sphere has only linear and quadratic UV
divergences, thus the finite piece, after appropriate
regularization, itself provides a
 measure of the number of degrees of freedom.

Moreover, I will show that for ${\cal N}=2$ SCFTs, there is an
explicit formula for $Z$ given the IR superconformal $R$-charge.
The function is extremized for the exact superconformal
$R$-charge. In examples, it appears that $Z$ is in fact always
minimized as a function of the $R$-charge associated to the
curvature couplings on the sphere.

The decrease in of the number of degrees of freedom of a field
theory along a renormalization group trajectory has been made
precise in two and four dimensions. The beautiful theorem of
Zamolodchikov \cite{Z} proves that in two dimensions the trace
anomaly, $c$, is strictly decreasing, and the rg flow is the
gradient flow of $c$.

In four dimensions, it is conjectured \cite{Cardy} that the
coefficient, $a$, of the Euler density appearing in the conformal
anomaly decreases along rg flows. Strong evidence for this was
found in presence of ${\cal N}=1$ supersymmetry \cite{BIWW}.
Moreover, an apparent counter-example \cite{ST} has recently been
removed \cite{GST}.

In theories with a gravity dual, the $S^3$ partition function is
identified with the exponential of  the euclidean Einstein action
in AdS$_4$. It diverges, and is regulated using a boundary counter
term \cite{HS1, HS2} (see also \cite{EJM, BK, KLS, SS}) leading to
a finite result for even dimensional AdS. The regularized action
turns out to be negative.

This gives some holographic evidence for a $Z$-theorem that $Z$ is
greater at the endpoint of a renormalization group flow
\cite{FGPW}, since the volume of the dual AdS must be less. If one
could prove that $Z$ was always minimized in determining the exact
$R$-symmetry, then the loss of flavor symmetries along rg flows
would give evidence for such a $Z$-theorem for ${\cal N}=2$
theories.

Many exact results involve the superconformal $R$-charge of ${\cal
N}=2$ conformal field theories in three dimensions. Just as in 4d
${\cal N}=1$ superconformal theories, it determines the dimensions
of chiral operators. Furthermore, as I will show, the partition
function on the sphere may also be determined given the exact
superconformal $R$-charge.

Until now, there was no known method to determine the
superconformal $R$-charge in three dimensions, beyond
considerations of symmetry. It can, of course, be calculated in
perturbation theory.

There is a further result known about the $R$-charges in 3d.
Barnes, Gorbatov, Intriligator and Wright showed \cite{BGIW}  that
the two point function of an $R$-current is minimized for the
superconformal one. This result gives another derivation of
$a$-maximization when applied to four dimensions. However, this
two point function receives quantum corrections, and cannot be
computed exactly in three dimensions.

The $S^3$ partition function that I define and calculate in this
paper is an explicit function of an $R$-charge. Extremizing it
gives an exact formula for the superconformal $R$-charge.

The recipe is as follows. Consider an ${\cal N}=2$ theory with $f$
abelian flavor symmetries. If $R_0$ is an $R$-charge, then so is
\begin{equation} R = R_0 + \sum_{j = 1}^f a_j F_j,\end{equation} where $F_j$ are the flavor
charges.

I will show that the partition function of the theory on $S^3$, as
a function of the $R$-multiplet used to couple to the background
geometry, is given by
\begin{equation} \label{partfunc} Z(R) = \int \prod_{Cartan} d u\
e^{i \pi \Tr u^2} {\det}_{Ad} \Big(\sinh(\pi u) \Big)
\prod_{\textrm{Chirals in rep } R_i} {\det}_{R_i} \Big(
e^{\ell(1-\Delta_i + i u)} \Big) ,\end{equation}
 where the $\Tr$ is the Chern-Simons form (normalized such that
 for $U(N)$ at level $k$, it is $k$ times the ordinary trace),
 $\Delta_i$ is the $R$-charge of the chiral multiplet under the
 $R$-symmetry, and the function $\ell$ is defined below.

 Suppose the $R$-symmetry can mix with a baryonic flavor symmetry associated to the conserved current $\star \tr F$.
 The $R$-charge may be expressed as $R = R_{matter} + \Delta_B B$, where $B$ is the topological (ie. monopole) charge.
 A factor of
 $\exp\left( 2 \pi \Delta_B \tr u\right)$ should then be included in
 \eqref{partfunc}, which is now a function of $\Delta_B$ as well.

The function \eqref{partfunc} is computed using localization, and
the resulting 1-loop determinants involve the function
\begin{equation}
\ell(z) = - z \log\left(1-e^{2\pi i z}\right) + \frac{i}{2}
\left(\pi z^2 + \frac{1}{\pi} \textrm{Li}_2(e^{2\pi i z}) \right)
- \frac{i \pi}{12},\end{equation} which satisfies $\del_z \ell(z)
= - \pi z \cot(\pi z)$.

 Setting $\del_{a_j} |Z|^2 = 0$ gives $f$ real equations for $f$
 unknowns, determining the exact superconformal $R$-charge in the
 infrared, up to possible discrete degeneracy.

As will be explained in section 2, an ${\cal N} = 2$ field theory
in three dimensions with an $R$-symmetry can be coupled to
curvature such that the 4 supercharges contained in $OSp(2|2)
\times SU(2)$ are preserved. That supergroup contains the $SO(4)
\cong SU(2) \times SU(2)$ isometries of $S^3$, but no conformal
transformations. The action which preserves this symmetry is very
similar to those appearing in \cite{Sen1, Sen2, Sen3, Romels} in
the related context of ${\cal N} = 1$ four dimensional theories on
$S^3 \times S^1$, and generalizes discussions for Yang-Mills
theories in \cite{Blau}. In general, there is a family of such
supersymmetric curvature couplings, parameterized by a choice of
$R$-charge. The linear combination of an
 $R$-charge with an abelian flavor charge is another $R$-charge.

The partition function may then be calculated using localization,
and is independent of the radius of the $S^3$. Thus the partition
function of the IR CFT, conformally coupled to curvature, can be
computed using localization of a UV theory. The latter is a function of the supersymmetric curvature couplings
that are parameterized by a choice of $R$-charge. It is equal to the partition function of the IR CFT conformally coupled to curvature
for the special choice of $R$-multiplet which sits in the superconformal algebra in
the IR. The result is calculated explicitly in section 3, as a
function of yet unknown IR $R$-charges. Here it must be assumed that
the $R$-symmetry does not mix with accidental $U(1)$'s.

In section 4, I consider deforming the theory by real mass
parameters, which preserve the same supersymmetry algebra. The
dependence of the supersymmetry transformations on the real masses
and the choice of $R$-multiplet used to couple the theory to
gravity is shown to be holomorphic. The structure of the
localization argument then implies that the partition function is
also a holomorphic function.

This relates the derivative with respect to the $R$-charge to the
one point function of an operator. As explained in section 5, this
operator may mix with the identity in the IR, but parity implies
that its VEV must be purely real in the conformal field theory. I
conclude that $|Z|^2$ is extremized when the $R$-symmetry is taken
to be the exact IR value. This determines the dimensions of all
chiral operators in the IR CFT.

In the last section, I check the proposal in several examples.

\end{section}

\begin{section}{Supersymmetric theories on $S^3$}

I begin by explaining how to put quantum field theories with 4
supercharges and an $R$-symmetry on $S^3$, while preserving
supersymmetry.

There is a canonical way of coupling conformal field theories to
curvature, defined by requiring Weyl invariance. Our ultimate aim
is to compute the partition function of an ${\cal N}=2$ SCFT on
$S^3$, conformally coupled to curvature.

One main result will be that this partition function can be
calculated exactly using localization in a UV Lagrangian
description. A prerequisite is the existence and uniqueness, in
the more general non-conformal context, of curvature couplings on
a round $S^3$ that preserve particular supersymmetries.

\subsection{Superconformal symmetries on $S^3$}

Recall that in three Euclidean dimensions, $\CN=2$ supersymmetry
implies two complex spinors.

In Lorentz signature, the superconformal group is $OSp(2|4)$,
containing the $SO(2)$ $R$-symmetry and $USp(4) \cong SO(3,2)$
conformal group. In Euclidean signature, the conformal group, in the notation of \cite{DFLV}, is $USp(2,2)$, the
real form of $USp(4)$ that is equivalent to $SO(4,1)$. Thus the
superconformal group in $\mathbb{R}^3$ is $OSp(2|2,2)$.

Denote the super(conformal) charges by $Q_A^i$, where $A=1, \dots,
4 $ is an $SO(4,1)$ spinor index  and $i=1,2$ is an $SO(2)$
$R$-symmetry index. 
The anti-commutator of the supersymmetries is given by
$$\{Q_A^i, Q_B^j\} = \delta^{ij} M_{AB} + i \omega_{AB} \epsilon_{ij}
R,$$ where $M_{AB} \in USp(2,2)$ and $R$ is the $U(1)$ superconformal
$R$-symmetry. One has that $[R, Q^i_A] = \epsilon^{i j} Q^j_A$.

Here, $\omega$ is the symplectic form of $USp(2,2)$,
$$\omega = \left(\begin{array}{cc}
\epsilon & 0\\ 0 &\epsilon\end{array} \right),$$ where 
$\epsilon = \left(\begin{array}{cc} 0 & 1\\ - 1 &
0\end{array}\right),$ is the anti-symmetric symbol on each $SU(2)$
factor.

On the sphere, the superconformal group is the same as in flat
space, however the interpretation of the bosonic $USp(2,2)$ is
different. In particular, the full $SU(2)_L \times SU(2)_r \cong
SO(4)$ subgroup\footnote{I reserve capital R for an $R$-symmetry.
$S^3$ is isomorphic to $SU(2)$, and these are the left and right
group actions.}  is realized as isometries of $S^3$, while only
the diagonal $SU(2)$ are rotations of $\mathbb{R}^3$ in the flat
space limit.

Localization of the path integral requires only a single global
supercharge. On $S^3$ there are four Killing spinors, satisfying
the Killing equation $\nabla_\mu \varepsilon = \gamma \varepsilon'$.

There is a basis of Killing spinors\footnote{My conventions for
spinors are collected in the appendix.} which further satisfy
$\varepsilon' \propto \varepsilon$. In particular, on $S^3$, one
has $\nabla_\mu \varepsilon = \pm \frac{i}{2 r} \gamma_\mu
\varepsilon$, where $r$ is the radius of the $S^3$. Following
\cite{KWY1}, choose such an $\varepsilon$, with the plus sign,
normalized such that $\varepsilon^\dag \varepsilon = 1$.

Take the holomorphic supersymmetry with spinor parameter
$\varepsilon$ that satisfies the homogeneous Killing spinor
equation to be $\delta = \frac{1}{\sqrt{2}}(Q^1_1 + i Q^2_1)$,
with $R$-charge $+1$. If $\delta$ is a symmetry of any Euclidean
theory on the sphere whose flat space limit is the analytic
continuation of a unitary Lorentzian theory, then $\delta^\dag =
\frac{1}{\sqrt{2}}(Q^1_2 - i Q^2_2)$ must also be a symmetry. In
Euclidean signature, $\delta^\dag$ now denotes not the Hermitian
conjugate, but an independent supercharge, defined by
$(\tilde{Q})_A^i  = \omega_{AB} \epsilon^{ij} Q^j_B$. It is easy
to check that $\delta^2 = (\delta^\dag)^2 = 0$.

Thus one must preserve
$$\{\delta, \delta^\dagger\} = M_{12} + R,$$ where
$M_{12}$ is  a rotation of $S^3$, and $R$ is the $R$-symmetry. The
particular rotation which appears is in the direction of the
vector field $v_\mu = \varepsilon^\dag \gamma_\mu \varepsilon$, as
defined in \cite{KWY1}. It can be viewed as translation along the
Hopf fiber of $S^1 \hookrightarrow S^3 \rightarrow S^2$.

More abstractly, there is an $OSp(2|2)_r \times SU(2)_L$ subgroup
of $OSp(2|2,2)$, containing the $SO(4) \cong SU(2)_L \times
SU(2)_r$ rotations of $S^3$, the  $R$-symmetry, and 4
supercharges. The supersymmetry $\delta$ is part of a doublet
under the $SU(2)_r$; it is a singlet of $SU(2)_L$.

The subgroup $OSp(2|2)$ does not contain any of the conformal
transformations. It is somewhat novel that the $R$ charge appears
on the right hand side of a supersymmetry algebra without
conformal generators. This also occurs for the supersymmetry group
$OSp(2|4)$ in Pestun's work \cite{Pestun} on  ${\cal N}=2$
theories on $S^4$.

 Regarding
$S^3$ as a Hopf fibration of $S^1$ over $S^2$, one $SU(2) \subset
SO(4)$ acts as rotations on the $S^2$ (together with the induced
action on the fiber), while the other contains a $U(1)$ subgroup
of the translations along the $S^1$ fiber. In the flat space
limit, the diagonal $SU(2)$ becomes the $SO(3)$ rotation group
that fixes a given point.

The $\delta$ used in localization is one of the fermionic
generators in $OSp(2|2)$. Note that this subgroup is preserved
under conjugation by $SO(4)$ rotations of the $S^3$. Thus $\delta$
and $\delta^\dag$, together with the $SO(4)$ isometries,  do not
generate the entire superconformal group.

Some of the theories considered in this work will be invariant
under parity. On $\mathbb{R}^3$, parity acts by reflection of one
coordinate, or, equivalently (up to an $SO(3)$ rotation) by
inversion through the origin.

This becomes a reflection through an equatorial $S^2$ on the
sphere, or, after an $SO(4)$ rotation, the antipodal map. Thus
parity acts by exchanging the two $SU(2)$ isometries.

Therefore, any theory invariant under $OSp(2|2)$ and this
$\mathbb{Z}_2$ inversion must in fact have the full superconformal
symmetry. The converse is of course false: there are
superconformal theories (invariant under $OSp(2|2,2)$ on the
sphere) that are not parity invariant. This includes most theories
with Chern-Simons terms.

\subsection{Supersymmetric curvature couplings}

I will now show explicitly how to couple an ${\cal N}=2$
Lagrangian field theory with an $R$ symmetry to the curvature of a
round $S^3$ such that $OSp(2|2)$ is preserved. In addition to
ordinary curvature couplings, proportional to $1/r^2$, there will
be terms in the action that depend on $1/r$, so calling them
curvature couplings is a slight misnomer.

In general, there will be a whole family of actions, differing
only in terms that vanish in the flat space limit, which preserve
$OSp(2|2)$. They will be parameterized by a choice of
$R$-symmetry. Recall that an $R$-symmetry together with any
combination of abelian flavor symmetries is again an $R$ symmetry.

The $R$-charges that appears on the right hand side of the algebra
will differ by abelian flavor charges. The couplings to curvature
are uniquely determined given that $R$-multiplet.

Consider a single chiral multiplet, coupled to an abelian gauge
field. Let $\Delta$ be the $R$-charge of the lowest component. The
supersymmetry transformations are
\begin{align}\label{mattersusy} \delta \phi& = 0\\ \delta \phi^\dag &= \psi^\dag \varepsilon \\
\delta \psi &= (-i \slashed{D} \phi - i \sigma \phi + \frac{\Delta}{r} \phi) \varepsilon\\
\delta \psi^\dag &= \varepsilon^T F^\dag\\
\delta F &= \varepsilon^T (-i \slashed{D} \psi + i \sigma\psi +
\frac{1}{r}(\frac{1}{2}-\Delta) \psi + i \lambda \phi)\\
\delta F^\dag &= 0,
\end{align} and
\begin{align} \delta A_\mu &= -\frac{i}{2} \lambda^\dag
\gamma_\mu \varepsilon \\
\delta\sigma &= -\frac{1}{2}\lambda^\dag \varepsilon\\
\delta \lambda &= \left(-\frac{1}{2} \gamma^{\mu \nu} F_{\mu \nu}
- D + i \gamma^\mu \del_\mu \sigma - \frac{1}{r} \sigma \right)
\varepsilon\\
\delta \lambda^\dag &= 0\\
\delta D &= \left( -\frac{i}{2} (D_\mu \lambda^\dag) \gamma^\mu +
\frac{1}{4 r} \lambda^\dag \right) \varepsilon.\end{align} Here
$r$ is the radius of $S^3$, $\gamma^{\mu \nu} =
\frac{1}{2}[\gamma^\mu, \gamma^\nu]$, and the gauge covariant
derivative is defined as $D_\mu = \del_\mu + i [A_\mu, \cdot ]$.

 When $\Delta = \frac{1}{2}$,
this is a symmetry of the classical action of an ${\cal N}=2$
chiral multiplet with canonical dimension, conformally coupled to
curvature .

As before, $\varepsilon$ is a Killing spinor satisfying
$\nabla_\mu \varepsilon = \frac{i}{2} \gamma_\mu \varepsilon$ and
$\varepsilon^\dag \varepsilon = 1$. Define a vector field $v_\mu =
\varepsilon^\dag \gamma_\mu \varepsilon = - \varepsilon^T
\gamma_\mu \varepsilon^*$, which satisfies $\nabla^\mu v_\mu = 0$,
as shown in \cite{KWY1}. Note that $\epsilon^\dag$ is also a left
invariant Killing spinor.

It is easy to check that these transformations satisfy the above
algebra,
\begin{align} \{\delta, \delta^\dag\} \phi &= - i (v^\mu D_\mu + \sigma) \phi + \Delta \phi\\
 \{\delta, \delta^\dag\} \psi &= - i (v^\mu D_\mu + \sigma) \psi + (\Delta-1) \psi\\
\{\delta, \delta^\dag\} F &= - i (v^\mu D_\mu + \sigma) F +
(\Delta-2) F,
\end{align}  and likewise for their conjugates. Note that the
terms involving $\sigma$ are simply a gauge transformation of the
fields, and do not give rise to a nontrivial charge in the
supersymmetry algebra.

Suppose one takes a non-conformal theory. To put it on the sphere,
one needs to specify how to couple it to curvature. If the theory
were conformal, those couplings are uniquely determined by
requiring Weyl invariance.

Instead, consider a theory which preserves $OSp(2|2)$. For any
choice of the $R$-charge (and thus $\Delta$ in the above
expressions), one can find a unique action on the sphere such that
the theory is invariant under that supersymmetry.

In particular, on the sphere, the matter action that preserves
$OSp(2|2)$ is
\begin{equation}\begin{split} S = \int \sqrt{g} \Big( D_\mu
\phi^\dag D^\mu \phi + i \psi^\dag \slashed{D} \psi + F^\dag F +
\phi^\dag \sigma^2 \phi + i \phi^\dag D \phi - i \psi^\dag \sigma
\psi + i \phi^\dag \lambda^\dag \psi - i \psi^\dag \lambda \phi \\
+ \frac{\Delta-\frac{1}{2}}{r} \psi^\dag \psi + \frac{2 i}{r}
(\Delta - \frac{1}{2}) \phi^\dag \sigma \phi +
\frac{\Delta(2-\Delta)}{r^2} \phi^\dag \phi
\Big).\end{split}\end{equation} Similar actions have appeared
before in the work of D. Sen \cite{Sen1, Sen2, Sen3} and, more
recently, R\"omelsberger \cite{Romels}, in the related context of
the (superconformal) index of four dimensional ${\cal N}=1$
theories on $S^3$.

Note that when $\Delta$ differs from $1/2$, the $\mathbb{Z}_2$
parity is broken by the new terms required to preserve
supersymmetry on the sphere. This is not too surprising,
considering that the choice of $OSp(2|2)_L$ already breaks
parity, as does the choice of Killing spinor.

The careful reader will also note that the factor of $i$ in the
term $\psi^\dag \sigma \psi$ and the sign of $\phi^\dag \sigma^2
\phi$ correct errors in the literature; moreover the auxiliary
field, $D$, is defined with a slightly unusual factor of $i$.

The supersymmetry transformation of the vector multiplet is
identical with that appearing in \cite{KWY1}, since those fields
cannot have anomalous dimensions ($\ast \Tr F$ is a conserved
current, and must have dimension 2 in the IR).

Therefore the unique supersymmetric Yang-Mills action on $S^3$ (preserving $OSp(2|2) \times SU(2)_L$) is
given by
\begin{equation}\label{YM}  \frac{1}{g_{YM}^2} \int \sqrt{g} \Tr\left( \frac{1}{2} F^{\mu \nu} F_{\mu \nu} + D_\mu \sigma D^\mu \sigma + D^2 + i \lambda^\dag \slashed{\nabla} \lambda + i [\lambda^\dag, \sigma] \lambda + \frac{2}{r} D \sigma  - \frac{1}{2 r} \lambda^\dag \lambda + \frac{1}{r^2} \sigma^2 \right).\end{equation}

Note that the curvature couplings of the scalars in the vector
multiplet already break parity, since $\sigma$ is a pseudo-scalar,
while $D$ is an ordinary scalar. There is a parity violating mass for the gauginos as well.

This avoids the following potential contradiction. QED with $N_f$
conjugate flavor pairs is parity invariant, yet has nontrivial
anomalous dimensions. Given that $OSp(2|2) \times SU(2)_L$ and
parity generate the entire superconformal algebra, if the
curvature couplings of pure YM had not broken parity, then neither
would those QED with flavors for the canonical choice of
dimensions. But that theory is not a CFT, so it is impossible.

Another important point is that the YM action \eqref{YM} is not
only invariant under supersymmetry, it is in fact $Q$-exact.

Therefore, by the standard localization argument, the partition
function does not depend on the dimensionless parameter $r
g_{YM}^2$. One can easily check that the $S^3$ partition function
computed using localization in the next section will in fact
always independent of the radius of the sphere.

The addition of a superpotential does not change the action of the
supercharges nor the value of the action on the localized
configuration space discussed in the next section. Therefore it
does not change the $S^3$ partition function.

Of course, superpotentials will typically break some flavor
symmetries, and thus will restrict the possible $R$-multiplets. In
other words, only a subset of the possible curvature couplings
discussed above will still preserve supersymmetry - namely those
associated to $R$-symmetries that are unbroken by the
superpotential.

Any theory with four supercharges and an $R$-symmetry possesses an
$R$-multiplet \cite{KS}. If there are abelian flavor symmetries,
then there is a family of $R$-multiplets related by improvement
terms. Such a theory may be put in curved space by first coupling
it to supergravity.
Those supergravity theories do depend on which $R$-multiplet is
gauged.

It would be very interesting to understand in that more abstract
language precisely what background of the fields in the
supergravity multiplet must be turned on for the round $S^3$, so
that in addition to the $SO(4)$ isometries, the global
$OSp(2|2)$ supersymmetry is preserved.

This would explain the origin of the somewhat mysterious $1/r$
couplings, give a general interpretation of the preservation of
supersymmetry on this space with Killing spinors, and would
generalize the result to theories with an $R$-multiplet without a
Lagrangian description.

\subsection{The essential logic}

In the UV, there is a way of coupling the theory to curvature on a
round $S^3$ such the $OSp(2|2)$ algebra is preserved. There is a
family of such theories on the sphere, parameterized by the
$R$-charge which appears in the algebra.

The partition function can then be computed using localization,
which renders the 1-loop approximation exact, and the result does
not depend on the radius of the sphere. In the next section, I
will exactly calculate this partition function, generalizing
\cite{KWY1} to situations with nontrivial anomalous dimensions.
The path integral will be a function of the choice of $R$-charge
that parameterizes the couplings to curvature on $S^3$.

One important subtlety is that the 1-loop determinants are UV
divergent.  I will regulate them using zeta function
regularization, as in \cite{KWY1}, \cite{DMP}. This is justified
{\it a posteriori} by the various checks of the result.

Moreover, it is reasonable that the 1-loop determinant for a
chiral multiplet charged under several gauge and flavor groups
depends only on the natural combination of the real scalars,
$\sum_a q_a \sigma_a$, where $q_a$ are the changes under the
Cartan factors. Strictly speaking, this is not guaranteed since
the chiral field is not gauge invariant.

In later sections, it will be shown that the dependence of the
1-loop determinant on the parameters $\Delta_j$ is related by
analytic continuation to the dependence on background flavor gauge
fields.

Therefore, assuming the regulator is such that the previous two
paragraphs apply, the 1-loop determinant for a single free chiral
multiplet would be enough to obtain the general result. It would
be much more satisfactory to determine the principle which selects
zeta function regularization.

The IR CFT can be conformally coupled to curvature. It preserves
the whole $OSp(2|2,2)$. Therefore, the UV theory
``$OSp(2|2)$-coupled'' to curvature on a large $S^3$ will be the
same as the IR theory conformally coupled to curvature exactly if
the $R$-charge is chosen to be the $R$-symmetry that sits in the
IR superconformal algebra.

Therefore, the partition function of the CFT on $S^3$ can be
computed using localization of a UV Lagrangian definition of the
theory, which is coupled to curvature such that the $OSp(2|2)$
which is preserved contains the exact superconformal $R$-charge of
the IR theory.

\end{section}

\begin{section}{Localizing the partition function}

Consider any theory which is invariant under the supersymmetry
(\ref{mattersusy}). Then the addition of $Q$-exact terms to
the action will not change the partition function by the standard
localization argument. Thus one can compute
$$Z = \lim_{t\rightarrow\infty} e^{-S - t\{\delta, P\}},$$ for any sufficiently regular, odd operator $P$ that is
 invariant under the bosonic symmetry $\delta^2$. In the case at hand, $\delta^2=0$, so the last is not a constraint.

The action for the vector multiplets, including their couplings to
curvature, is identical to that appearing in \cite{KWY1}, since
those fields cannot have anomalous dimension. Thus the arguments
of Kapustin-Willett-Yaakov may be carried over directly.

In brief, \cite{KWY1} takes $P = \Tr\left( (\delta \lambda)^\dag
\lambda \right)$, where the $\delta \lambda$ is understood to be
stripped of the Killing spinor $\varepsilon$. The bosonic part of the
localizing action $S_{loc} = t \{\delta, P\}$ is positive
definite, and vanishes exactly when $\delta \lambda = 0$.

The only solutions are that $D = -\sigma/r$ is a constant
\footnote{I keep the radius of the sphere explicit; it is set to 1
in \cite{KWY1}.} and all other fields are set to zero. The path
integral reduces to a matrix model, that is, a finite dimensional
integral over the constant VEVs of $\sigma$.

The Yang-Mills action coupled to curvature such that $OSp(2|2)$
is preserved is $Q$-exact, and indeed restricts to zero on the
localized space of field configurations. In fact, the only
non-vanishing piece of the action is from the supersymmetric
Chern-Simons term,
$$\frac{i k}{4\pi} \int_{S^3} 2 \Tr (D \sigma) = i \pi k r^2
\Tr(\sigma^2),$$ where one uses the fact that the volume of $S^3$
is $2\pi^2 r^3$.

Turning now to the matter sector, a particularly natural choice,
as in \cite{KWY1}, is to take $$P = (\delta \psi)^\dag \psi +
\psi^\dag (\delta \psi^\dag)^\dag,$$ where in this expression,
$\delta \psi$ is understood to not include the $\varepsilon$.  This
ensures that the bosonic part of the $Q$-exact localizing term in
action, $S_{loc}$, is positive, and vanishes on supersymmetric
configurations. On $S^3$, this implies that all of the fields are
localized to zero with the exception of the scalars in the vector
multiplets. Integrating over their constant VEVs results in a
finite dimensional matrix integral.

The localizing Lagrangian, ${\cal L}_{loc} = \{\delta, P\}$, is
\begin{equation} \del_\mu \phi^\dag \del^\mu \phi + \phi^\dag
\sigma_0^2 \phi + \frac{2 i}{r} (1-\Delta) \phi^\dag v^\mu
\del_\mu \phi + \frac{\Delta^2}{r^2} \phi^\dag \phi +  F^\dag F +
\psi^\dag \left( i \slashed\nabla - i \sigma_0 + \frac{1}{2} +
(1-\Delta) \slashed{v} \right) \psi,\end{equation} up to total
derivatives. It breaks the right $SU(2)$ isometry down to $U(1)$.

The 1-loop determinant from the bosonic fields is $\det^{-1}
(D_{bos})$, where $$D_{bos} = - \nabla^2 + 2 i \frac{1-\Delta}{r}
v^\mu \del_\mu + \frac{\Delta^2}{r^2} + \sigma_0^2 = \frac{1}{r^2}
\left( - \ell_j \ell^j + 2 i (1-\Delta) \ell_3 + \Delta^2 +
\sigma_0^2 r^2\right),$$ in terms of the left invariant vector
fields defined in the appendix. After expanding in terms of
angular momentum modes on the $S^3$, one finds that
\begin{equation} r^{-2(\ell+1)} {\det}_{\ell/2} (D_{bos}) =
\prod_{m=-\ell/2}^{\ell/2} \left( \ell(\ell+2) - 4 m (1-\Delta) +
\Delta^2 + \sigma_0^2 r^2\right).\end{equation}

The fermionic action is also quadratic, and one needs to compute
the determinant of the operator $$r D_{ferm} = i \gamma^j \ell_j -
1 - i r \sigma_0 + (1-\Delta) \slashed{v} = - 4 \vec{S} \cdot
\vec{L} + 2 (1-\Delta) S_3 - 1 - i r \sigma_0.$$ Up to a change in
the
coefficients, 
 this is
identical to the 1-loop determinant in the case of fields with
canonical dimensions studied by \cite{KWY1}. They found that
\begin{equation} \begin{split}
{\det}_{\ell/2} \left( 2 \alpha \vec{L} \cdot \vec{S} + 2 \beta
S_3 + \gamma \right) = \left(\alpha \frac{\ell}{2} + \beta +
\gamma\right) \left(\alpha \frac{\ell}{2} -\beta + \gamma\right)
\times \\ \prod_{m=-\ell/2}^{\ell/2-1} \left(-\frac{\ell}{2}
(\frac{\ell}{2} + 1) \alpha^2 - (2m +1) \alpha \beta - \alpha
\gamma - \beta^2 + \gamma^2\right).\end{split}\end{equation}

Therefore, in this case,
$${\det}_{\ell/2} (D_{ferm}) = r^{-2(\ell+1)} \big(-\ell-\Delta-i \sigma_0 r
\big) \big(-\ell - 2 +\Delta - i \sigma_0 r\big)
\prod_{m=-\ell/2}^{\ell/2-1} \big( -\ell(\ell+2)+4(1-\Delta) m -
\Delta^2 - \sigma_0^2 r^2\big).$$

The matter determinants mostly cancel between bosons and fermions,
leaving the following infinite product.\footnote{See also
\cite{HHL} that appeared shortly after the first version of this
paper,
  which derived the same result also using localization of non-conformal $R$-symmetric ${\cal N}=2$
 theories on $S^3$.}
\begin{equation}\label{matterdet} Z_{1-loop} = \prod_{n=1}^\infty \left(\frac{n+1+i u - \Delta}{n-1-i u + \Delta}\right)^n,\end{equation}
 where I defined $u=\sigma_0 r$. 
 
Define $z = 1 - \Delta + i u$, and let $\ell(z) = \log
Z_{1-loop}$. Then $$\del_z \ell(z) = \sum_{n=1}^\infty \left(
\frac{n}{n+z} + \frac{n}{n-z}\right),$$ which has a linear
divergence. Regulating using zeta functions, one finds that
$$\del_z \ell(z) = \frac{\del}{\del s}\Big{|}_{s=0} \Big(
\zeta_H(s-1,-z)+z\zeta_H(s, -z)-\zeta_H(s-1,z) +
z\zeta_H(s,z)\Big),$$ where $\zeta_H$ is the Hurwitz zeta
function.

This results in $$\del_z \ell(z) = - \pi z \cot(\pi z),$$ which
can be integrated to give \begin{equation} \ell(z) = - z
\log\left(1-e^{2\pi i z}\right) + \frac{i}{2} \left(\pi z^2 +
\frac{1}{\pi} \textrm{Li}_2(e^{2\pi i z}) \right) - \frac{i
\pi}{12}.\end{equation} This is not manifestly real when $z$ is
real, but it is clear from the original definition that it will
be.

\end{section}

\begin{section}{Deformation by real masses and a mysterious
holomorphy}

One initially surprising feature of the matter one loop
determinant \eqref{matterdet} is that it is identical to the
result for $\Delta = 1/2$ found in \cite{KWY1} up to the
replacement $u \rightarrow u + i (\Delta - 1/2)$. In particular,
the dependence on the variable $z$ is holomorphic.

This can be explained simply by the fact that the supersymmetry
transformation, $\delta$, itself depends holomorphically on $z$.
This holomorphy was not manifest in the above calculation of the
1-loop determinant, since the operator $P$ depended
anti-holomorphically on $z$.

However, the final answer does not depend on the choice of $P$,
and only depends on $z$ due to the dependence of $\delta$.
Therefore it must be holomorphic. Note that a fixed $P$,
independent of $z$ could have been chosen in this case since
$\delta^2=0$, so there is no $z$ dependent constraint on $P$.

So far, this looks like a curious feature of the matter 1-loop
determinant of a chiral multiplet charged under an abelian gauge
field. However, it has more general implications.

Recall that the supersymmetry preserving couplings of the QFT to
the gravity background of $S^3$ are parameterized by the space of
abelian flavor symmetries. Coupling all of those flavor symmetries
to background ${\cal N}=2$ vector multiplets, one may turn on real
mass parameters, $m_j$, that are constant VEVs for the real
scalars in those background multiplets.

I will now explain that the $S^3$ partition function depends
holomorphically on the parameters $z_j = - a_j + i r m_j$, where
$j$ runs over the Cartan of the flavor symmetry group and the
chosen $R$-charge is $R_0 + \sum a_j F_j$. Here $F_j$ denote the
flavor charges, and $R_0$ is some $R$-charge.
It would be extremely interesting to understand the origin of this
holomorphy - it will remain an empirical observation in this
paper.

In flat space, the real mass deformation is defined by turning on
a constant $\theta\bar\theta$ component (denoted by $\sigma$) of a
background abelian vector field. On $S^3$, the supersymmetry
transformations are modified, as we saw above in the matter
sector. In the case of the vector multiplet, the supersymmetry
preserving configuration on $S^3$ is given by a constant $D =
-\sigma/r$, where $r$ is the radius of $S^3$, and $D$ is the
auxiliary, $\theta^2 \bar\theta^2$ component \cite{KWY1}.
Obviously, this reproduces the standard real mass in the flat
space limit.

Alternatively, one can use the standard definition of a real mass
deformation, and include extra terms in the curvature couplings
needed to supersymmetrize the theory on the sphere.

As observed in \cite{KWY2}, the supersymmetry transformation
written above is still a symmetry of the theory after such a real
mass has been turned on, simply by setting $\sigma$ to be the real
mass parameter, and taking the vector multiplet to be
non-dynamical. This modifies the supersymmetries relative to those
of the theory without real masses.

Moreover, the real mass parameter appears in the anti-commutator
of supercharges; it is a central charge of the algebra,
$$\{\delta, \delta^\dag\} = M_{12} + \frac{1}{r}R_{UV} + (\frac{a_j}{r} - i m_j) F_j.$$
Further
note that the supersymmetry transformation itself depends
holomorphically on $a-im$.

The real mass must break conformal invariance, however it
preserves $OSp(2|2)$ if one also turns on $D  = -\sigma/r$. I
will refer to that susy preserving deformation as the real mass on
$S^3$. In the flat space limit, it is just the usual real mass.

For example, consider a chiral multiplet with canonical dimension
$1/2$. Then invariance under supersymmetry requires that the
action on the sphere be given by
$$S = \int \sqrt{g} \left( g^{\mu \nu} \del_\mu \phi^\dag
\del_\nu \phi + i \psi^\dag \slashed\nabla \psi + F^\dag F + i m
\psi^\dag \psi + (m^2 - \frac{i m}{r} + \frac{3}{4 r^2}) \phi^\dag
\phi \right) .$$

The terms $i m \psi^\dag \psi - m^2 \phi^\dag \phi$ are the usual
real mass terms in Euclidean signature. The term $\frac{3}{4 r^2}
\phi^\dag \phi$ is the standard conformal coupling of a free
scalar. The term $- i \frac{m}{r} \phi^\dag \phi$ can be
understood as arising because a constant value of $D =
-\frac{m}{r}$ must be turned on in the background vector multiplet
to preserve supersymmetry on $S^3$.

The partition function on the sphere can be computed using
localization. It depends on the real masses, as well as the
$R$-multiplet used to couple the theory to curvature. The
mechanism of localization implies that the only dependence on
$m_j$ and $\Delta_j$ is through the supersymmetry transformation,
$\delta$, and the value of the action on the localized space of
field configurations.

The terms $\int D \sigma$ appearing in ${\cal N}=2$ Chern-Simons
are the only ones in a YM-CS-matter action that are non-zero on
the space of configurations for which the fermion variations
vanish. These have no dependence on $m_j$ or $\Delta_j$.

Therefore the full $S^3$ partition function depends on $m_j$ and
$\Delta_j$ only through the supercharge, and it inherits the
holomorphy, $Z(m_j, \Delta_j) = Z(\Delta_j - i r m_j)$.

It is natural to assume that this holomorphy also applies to
baryonic flavor symmetries. Their associated real masses are
Fayet-Iliopoulos parameters, appearing in the action as $-
\frac{i}{\pi} \xi \Tr(D)$.

After localization, this results in a term $-
\frac{i}{\pi}\int_{S^3} \xi \Tr(-\sigma/r) = 2\pi i  \xi
\Tr(\sigma) r^2$, as in \cite{KWY2}. Therefore, I conjecture that
if the $R$-symmetry mixes with a baryonic $U(1)$, then the
partition function \eqref{partfunc} is modified by the inclusion
of a factor $$e^{2 \pi \Delta_B \Tr(u)}$$ in the integrand. This
is consistent with the equivalence of the baryonic $U(1)$ and the
gauged matter current in abelian gauge theories with a non-trivial
Chern-Simons level (so that it is really $j_{matter} + k \star F$
that is gauged).

\end{section}

\begin{section}{$Z$-extremization}

The holomorphy of the partition function implies that $\del_\Delta
Z = \frac{i}{r} \del_m Z$. At the conformal point, $\frac{1}{Z}
\del_m Z\big{|}_{m=0, \Delta = \Delta_{\textrm{IR}}}$ is the
1-point function of an operator in a CFT on $S^3$.

Thus, by conformal invariance, it vanishes unless the operator
contains the identity. In general, the operator associated to a
real mass, as defined in the UV theory, may indeed mix with the
identity.

First consider the case that the CFT is parity symmetric, and it
is not spontaneously broken. The real mass term is parity odd.

Therefore, its VEV must vanish, since parity isn't spontaneously
broken, by assumption. This argument does not appear to utilize
conformal invariance, but recall that parity, together with
$OSp(2|2) \times SU(2)_L$,
 generates the entire superconformal algebra.

 In parity invariant theories, $Z$ is a real number, and the
 requirements that $\del_{\Delta_j} Z(\Delta_j) = 0$ are exactly the right number of equations to determine all
 of the $\Delta_j$ (up to a possible discrete degeneracy).

In general, the CFT will not preserve parity. Then the $S^3$
partition function will be complex, and parity acts on it by
complex conjugation.

In a non-conformal field theory without parity, the VEVs of parity
violating operators may be non-zero. However in a CFT, only the
identity operator has a non-vanishing VEV.

The identity operator is parity invariant, thus its VEV must be a
real number, $\textrm{Im}\left(\frac{1}{Z} \del_{m_j} Z \right) =
0$, where the partition function is evaluated at the conformal
point.

Therefore, \begin{equation} \del_{\Delta_j} |Z|^2 =
0.\end{equation} This is the main result. Given the calculation of
section 3, it is an explicit formula, in terms of the UV data of
gauge groups, matter representations and Chern-Simons levels, that
determines the exact $R$-charge in the superconformal algebra of
the IR theory.

It may seem somewhat unusual that the operator which couples to
the real mass at linear order mixes with the identity operator in
parity violating theories. The effect appears already at 1-loop in
perturbation theory, when computing the VEV of $\psi^\dag \psi$ in
the UV theory.

For example, in a Chern-Simons theory at level $k$ with some
charged matter, there is a 1-loop diagram involving $\psi^\dag
\psi$, the two fermion-two boson vertex, and a $\phi$ loop, at
order $1/k$. It is quadratically divergent, and in flat space one
usually throws it away.

However, on $S^3$, the zeta function regularization instructs one
to keep a finite piece.

It would also be interesting to look at the second derivative,
which is related to the two point function. Its positivity would
imply that $|Z|^2$ was always minimized, which is indeed seen in
examples. 

The two point function is integrated over the entire $S^3$, so
contact terms must be taken into account. Similarly, there may be
explicit $m^2$ couplings in the action. I leave this question  and
that of the interpretation of multiple extrema of $|Z|^2$ for
future work.

\end{section}

\begin{section}{Examples}

I will now check the above prescription in several examples, which
for simplicity involve only a single integral after localization.

\subsection{${\cal N}=4$ vector multiplet}

The ${\cal N}=4$ vector multiplet contains,  in ${\cal N}=2$
language, an adjoint vector superfield and an chiral superfield,
$\Phi$. The latter has dimension 1, rather than the canonical
dimension $\half$.

Such theories were studied in \cite{KWY2}. To obtain answers in
agreement with 3d mirror symmetry, those authors set the 1-loop
determinant arising from that chiral field in the ${\cal N}=4$
vector multiplet to 1, a constant independent of the variables in
the matrix integral.

This was justified by noting that an F-term mass for that adjoint
chiral, $\int d^2 \theta \Phi^2$, is in fact Q-exact. Thus $\Phi$
may be localized to zero, independently of the rest of the theory,
and its contribution to the 1-loop determinant must be trivial.

The formula (\eqref{matterdet}) implies that 1-loop determinant of
an adjoint chiral multiplet with dimension 1 is
$$\prod_{i < j} \prod_{n=1}^\infty \left( \frac{n + i (u_i -
u_j)}{n - i (u_i - u_j)} \right)^n \left( \frac{n + i (u_j -
u_i)}{n - i (u_j - u_i)} \right)^n = 1,$$ as predicted.

\subsection{SQED}

Consider ${\cal N}=2$ QED with $N_f$ conjugate flavor pairs, $Q$
and $\tilde Q$. This theory has a $U(1)$ flavor symmetry under
which the flavors all have the same charge. There is also a
topological $U(1)_B$, which cannot mix with the $R$-charge in this
charge conjugation invariant theory, and an $SU(N_f)$ flavor
symmetry, which cannot mix with the abelian $R$-symmetry.

The partition function of the theory coupled to curvature on $S^3$
using the $R$-multiplet under which the flavors have charge
$\Delta$ is given by
\begin{equation}
Z = \int_{-\infty}^\infty d u \ \exp\Big( N_f( \ell(1-\Delta  + i
u) + \ell(1-\Delta - i u))\Big).
\end{equation}
SQED is parity invariant, and this is a real valued function.

Setting $\del_\Delta Z = 0$ determines the superconformal value of
$\Delta$. In particular, for $N_f=1$, one can check numerically that
$\Delta=1/3$. An analytic proof will be given below.

This is precisely the prediction of 3d mirror symmetry, which
relates SQED with one flavor to the $X Y Z$ model \cite{AHISS}.
This Wess-Zumino model has three chiral fields, and a
superpotential, ${\cal W} = X Y Z$ with an $S_3$ discrete
symmetry.

 The gauge invariant operator
$\tilde Q Q$ of SQED therefore has dimension $2/3$, since it
corresponds to the chiral operator $X$.

The three dimensional mirror symmetry relating ${\cal N}=4$ SQED with one hypermultiplet flavor and the free theory of a twisted hyper implies the equality of their sphere partition functions. This was confirmed in \cite{KWY2}, and generalized to the case of ${\cal N}=2$ preserving real masses in \cite{HLP}. Using the holomorphy, the ${\cal N}=2$ relation with the $X Y Z$ model can also be seen analytically as follows.\footnote{I would like to thank Davide Gaiotto for suggesting this argument.}

The partition function of  ${\cal N}=4$ SQED with one flavor
deformed by the ${\cal N}=2$ real mass is given by
$$Z^{SQED}_{{\cal N}=4}(m) = \int_{-\infty}^\infty du\
e^{\ell(1/2+ i m + i u) + \ell(1/2 + im - i u) + \ell( - 2 i
m)}.$$ The first two terms are from the conjugate pair of
fundamental chirals with dimension $1/2$, and the third term
derives from the adjoint chiral with dimension 1 in the ${\cal
N}=4$ vector multiplet. The last term was neglected in the
analysis of \cite{HLP}, resulting in a slight mismatch that they
remark on.\footnote{See also version 2 of \cite{HLP} that appeared
shortly after this paper.} There is a superpotential ${\cal W} =
\tilde{Q} \Phi Q$, so the real mass is associated to a flavor
$U(1)$ under which $\Phi$ has charge $-2$ when $Q$ and $\tilde{Q}$
have charge 1.

The statement of 3d mirror symmetry is then
$Z^{SQED}_{{\cal N}=4} (m) = e^{2 \ell(1/2 - i m)}$, which corrects the result of \cite{HLP} by the factor of $e^{\ell(1-2i m)}$ on the left hand side. To relate this to ${\cal N}=2$ SQED, which does not have the adjoint chiral multiplet, one must divide the whole equation by its contribution.

Therefore, $$\int_{-\infty}^\infty d u\ e^{\ell(1-\Delta + i u) + \ell(1-\Delta - i u)} = e^{2\ell(\Delta) - \ell(2\Delta - 1)},$$ using the holomorphy to set $m = i (\Delta-1/2)$. The expression on the right hand side is exactly the sphere partition function of the $X Y Z$ model with $R$-charges taken to be $1-\Delta$, $2 \Delta$ and $2 \Delta$. It is trivial to check that it is minimized by the symmetric choice, $\Delta = 1/3$.

\subsection{Abelian Chern-Simons}

Adding an ${\cal N}=2$ Chern-Simons term to the previous example
breaks parity. A term $e^{i \pi k u^2}$ is introduced into the
matrix integral. The superconformal $R$-charge is determined by
setting $\del_\Delta |Z|^2$
to zero. 

The partition function can be evaluated in the large $k$ limit to
give
\begin{equation}\begin{split} \int_{-\infty}^\infty d u \ e^{i \pi
k u^2} e^{ N_f( \ell(1-\Delta + i u) + \ell(1-\Delta - i u))} =
\frac{1}{2^{N_f}} \sqrt{\frac{i}{k}} \textrm{\Huge[} 1 + \frac{N_f
\pi^2 a^2}{2} + \frac{i \pi N_f}{4 k} \left( 1 + 4 a + \pi^2 a^2 +
\frac{\pi^2 N_f a^2}{2}
\right) \\
+ \frac{\pi^2 N_f}{16 k^2} \left( 1 - 8 a - \frac{3}{2} N_f - 12 a
N_f - 24 a^2 N_f - 4 a^2 \pi^2 - \frac{7}{2} a^2 N_f \pi^2 -
\frac{3}{4} a^2 N_f^2 \pi^2 \right) + {\cal
O}\left(\frac{1}{k^3}\right) \textrm{\Huge]},\end{split}
\end{equation} where $\Delta = \frac{1}{2} + a$.

The dimensions of the fundamentals in the IR were calculated
perturbatively in the Chern-Simons-matter theory by Gaiotto and
Yin \cite{GY},
$$\Delta = \frac{1}{2} - \frac{b_0}{4 k^2} + {\cal
O}\left(\frac{1}{k^4}\right),$$ where $b_0 = \frac{2}{\dim R}\big(
\tr_R (T^a T^b) \tr_R (T^a T^b) + \tr_R (T^a T^b T^a T^b) \big),$
for matter in a  representation, $R \oplus \bar{R}$, of the gauge
group with Lie algebra generated by $T^a$, normalized\footnote{This differs
slightly from the normalization in \cite{GY}.} such that
${\tr}_{fund}( T^a T^b) =  \delta^{a b}$.

The $R$-charges determined by minimizing $|Z|^2$ are given by
$$\Delta = 1/2 - \frac{N_f + 1}{2 k^2} + {\cal
O}(\frac{1}{k^4}),$$ in perfect agreement with the perturbative
field theory calculation, which gives $b_0 = 2 (N_f+1)$.

\subsection{$SU(2)$ Chern-Simons}

The integral over the Cartan of $SU(2)$ again involves only a
 single integration variable. The vandermonde and 1-loop
determinant for the gauge sector are now non-trivial.

Consider $SU(2)$ Chern-Simons theory at level $k$ and $N_f$ chiral
multiplets in the self-conjugate $2$ representation. The
$R$-symmetry may mix with the $U(1)$ flavor symmetry under which
all of the matter has the same charge.

The partition function on the sphere is given by
$$Z = \int_{-\infty}^\infty d u\ \sinh^2(2 \pi u) e^{2 i \pi k u^2} e^{N_f(\ell(1-\Delta+ i u)+\ell(1-\Delta-i u))},$$
 up to an overall $\Delta$ independent constant.

Calculating the large $k$ asymptotics of $Z$, and solving the
equation $\textrm{Re}\left(\del_\Delta \log Z \right) = 0$,
determines the infrared dimensions to be $$\Delta = 1/2 -
\frac{3}{8 k^2} (N_f-1) + {\cal O}(\frac{1}{k^4}),$$ again
agreeing with the perturbative computation, for which $b_0 =
\frac{3}{2}(N_f - 1)$. In this case, the representation $R$ is
$\frac{N_f}{2}$ copies of the fundamental of $SU(2)$, in the
notation of \cite{GY} in which the matter is in the $R \oplus
\bar{R}$.

For $g$ adjoint chiral multiplets, the method of $Z$-extremization
gives $$\Delta = \frac{1}{2} - \frac{4}{k^2} (g+1).$$ This exactly
matches the two-loop calculation in appendix (D.1) of \cite{GY}.

\end{section}

\subsection*{Acknowledgments}
I would like to thank Thomas Dumitrescu, Guido Festuccia, Davide
Gaiotto, Anton Kapustin, Zohar Komargodski, Juan Maldacena, Nathan
Seiberg, David Shih, Yuji Tachikawa, Xi Yin, and Edward Witten for
stimulating and insightful discussions. I am supported in part by
DOE grant DE-FG02-90ER40542.

\begin{appendix}

\section{Conventions for spinors and derivatives on $S^3$}

I use the following conventions, based on \cite{KWY1}.

On $S^3$, spinors have 2 complex components, and transform in the
fundamental of $SU(2) = Spin(3)$.

Regarding $S^3$ as the $SU(2)$ group manifold, one can find an
orthonormal triplet of vector fields that are invariant under the
left (resp. right) action of $SU(2)$, denoted by $\ell_\mu^i$
(resp. $\scripty{r}_\mu^i$), where $i = 1,2,3$ is a frame index.

The Laplacian on the sphere of radius $r$ can be expanded as $r^2
\nabla^2 = \sum (\scripty{r}^i)^2 = \sum (\ell^i)^2$. Defining $L_i = - \frac{i}{2} \ell_i$, one can check that $[L_i, L_j] = i \epsilon_{i j k} L_k$.

One can use the left invariant vector fields to define a vielbein
as the dual 1-forms, $e^i_\mu$. The gamma matrices can be written
using frame indices,  $\gamma_\mu = e^k_\mu \gamma_k$, where the
$\gamma_k$ are just the Pauli matrices, satisfying $[\gamma_i,
\gamma_j] = 2 i \epsilon_{i j k} \gamma^k$. Let $S^i = \frac{1}{2}
\gamma^i$.

In the left invariant frame, the spin connection is given by
$\omega_{i j} = \epsilon_{i j k} e^k$, and the spinor covariant
derivative is $\nabla_\mu = \del_\mu + \frac{i}{2} e^k_\mu
\gamma_k$.

Half of the 4 Killing spinors on the sphere are constant in this
frame, satisfying $\nabla_\mu \varepsilon_a = \frac{i}{2} e^k_\mu
\gamma_k \varepsilon_a = \frac{i}{2} \gamma_\mu \varepsilon_a$.
These 2 Killing spinors are invariant under the left $SU(2)$ by
construction, and transform in a doublet of the right $SU(2)$
(that is the $a$ index).

Pick one of the left invariant Killing spinors, and define a left
invariant vector field $v_\mu = \varepsilon^\dag \gamma_\mu
\varepsilon$, which can be identified as $\ell^3$. Therefore
$\slashed{v} = \gamma^3 = 2 S^3$.

The other pair of Killing spinors satisfy $\nabla_\mu \eta_b = -
\frac{i}{2} \gamma_\mu \eta_b$, and transform in a doublet of the
left $SU(2)$. They are constant in the right invariant frame
constructed from the \nolinebreak $\scripty{r}^i_\mu$.

\end{appendix}

\end{document}